\documentclass[conference]{IEEEtran}
\IEEEoverridecommandlockouts
\usepackage{bbm}

\usepackage{cite}
\usepackage{amsmath,amssymb,amsfonts}
\usepackage{graphicx}
\usepackage{textcomp}
\usepackage{xcolor}
\usepackage{cite}
\usepackage{dblfloatfix}
\usepackage{subcaption}
\usepackage{overpic}
\usepackage{amsmath,amssymb,amsfonts}

\usepackage{graphicx}
\usepackage{textcomp}
\usepackage{xcolor}
\usepackage{float}
\usepackage{amsthm}
\usepackage{graphicx}
\usepackage{epstopdf}
\usepackage{amsmath,bm}
\usepackage{amsfonts}
\usepackage{amssymb}
\usepackage{color}
\usepackage{multirow}
\usepackage{multicol}
\usepackage{soul,xcolor}
\usepackage{algorithm}
\usepackage{algpseudocode}

\usepackage{comment}

\theoremstyle{plain}

\newcommand{\vect}[1]{\mathbf{#1}}

\def\diag{\mathrm{diag}}

\def\tr{\mathrm{tr}}

\def\Htran{\mbox{\tiny $\mathrm{H}$}}
\def\Ttran{\mbox{\tiny $\mathrm{T}$}}
\def\CN{\mathcal{N}_{\mathbb{C}}}

\def\tr{\mathrm{tr}}

\def\BibTeX{{\rm B\kern-.05em{\sc i\kern-.025em b}\kern-.08em
    T\kern-.1667em\lower.7ex\hbox{E}\kern-.125emX}}
\begin{document}

\title{Distributed Two-Phase Processing for Modular XL-MIMO with Wireless Fronthaul under Hardware Impairments
\thanks{This work was carried out within the scope of the project 122C149 – Intelligent End-to-End Design of Energy-Efficient and Hardware Impairments-Aware Cell-Free Massive MIMO for Beyond 5G. \"O. T. Demir was supported by the 2232-B International Fellowship for Early Stage Researchers Programme funded by the Scientific and Technological Research Council of Türkiye (TÜBİTAK).}
}

\author{\IEEEauthorblockN{\"Ozlem Tu\u{g}fe Demir}
\IEEEauthorblockA{\textit{Department of Electrical and Electronics Engineering} \\
\textit{Bilkent University}\\
Ankara, Turkiye \\
E-mail: ozlemtugfedemir@bilkent.edu.tr}
\vspace{-10mm}
}

\maketitle

\begin{abstract}
Modular extremely large-scale MIMO (XL-MIMO) architectures combined with wireless fronthaul provide a scalable alternative to monolithic arrays, but their performance is sensitive to hardware impairments and resource allocation strategies. 
In this paper, we consider a distributed two-phase processing framework for modular XL-MIMO systems employing amplify-and-forward wireless fronthaul under practical hardware constraints. 
We jointly model access-side and fronthaul-side distortions and formulate a weighted minimum mean-square error (WMMSE)-based optimization problem that maximizes the uplink sum spectral efficiency (SE) by jointly adjusting UE transmit powers and fronthaul amplification levels. 
The resulting algorithm alternates between distortion-aware receiver design and convex power-control updates. 
Numerical results demonstrate that the proposed joint optimization significantly improves spectral efficiency compared to fixed transmission strategies, particularly when the CPU has a moderate number of antennas, while also quantifying the relative impact of access and fronthaul impairments.
\end{abstract}

        \vspace{-2mm}

\section{Introduction}

Extremely large-scale multiple-input multiple-output (XL-MIMO), also known as extremely large-scale antenna arrays (ELAAs), has recently attracted significant attention due to its large beamforming gains, high spatial resolution, and favorable near-field propagation characteristics \cite{hu2018beyond,bjornson2024towards}. 
When the physical aperture grows, spherical wavefront propagation improves user separability and spatial multiplexing capabilities. 
However, coordinating very large monolithic arrays introduces substantial hardware complexity, control signaling overhead, and fronthaul requirements.

Modular array architectures provide a scalable alternative that preserves the large-aperture benefits of XL-MIMO while reducing implementation complexity \cite{kosasih2025near,li2023modular}. 
Instead of a fully co-located array, multiple smaller subarrays are distributed over large surfaces, enabling local processing and improved hardware scalability while maintaining favorable near-field effects.

To further reduce deployment costs, wireless fronthaul has emerged as an attractive replacement for expensive fiber infrastructure and has recently been studied in cell-free massive MIMO systems \cite{demirhan_wireless_fronthaul,ibrahim_wireless_fronthaul,ozan_wireless_fronthaul}. 
While prior works mainly consider digital fronthaul, amplify-and-forward (AF) wireless fronthauling has been shown to approach centralized uplink performance with significantly reduced infrastructure cost \cite{demir2025cell}.

In this paper, we extend AF-based wireless fronthaul to modular XL-MIMO systems, where coordinated subarrays serve multiple user equipments (UEs) over shared time--frequency resources and communicate with a central processing unit (CPU). 
Different from \cite{demir2025cell}, we formulate a weighted minimum mean-square error (WMMSE) based sum-rate maximization problem that jointly optimizes UE uplink powers and fronthaul amplification levels. 
We further incorporate hardware impairments at both the access and fronthaul sides and quantify their end-to-end impact, revealing when adaptive joint optimization provides substantial gains over fixed transmission strategies.

\vspace{-2mm}
\section{System Model}

We consider a \emph{modular XL-MIMO} architecture composed of $L$ distributed subarrays. 
Each subarray is equipped with $N$ antennas for uplink reception and is connected to a dedicated baseband unit (BBU) that performs local signal processing. 
The same $N$ antennas are also used to forward the received uplink signals to a CPU via a wireless fronthaul link. 
The CPU coordinates all BBUs and enables joint processing across the entire XL-MIMO system. 
It is equipped with $M$ antennas to receive the wireless fronthaul transmissions from the subarrays.

The subarrays may be deployed at different locations within a large building or across nearby buildings, forming a physically distributed yet logically unified array. 
The CPU is located in close proximity to the subarrays, enabling the construction of a modular XL-MIMO system with scalable aperture and processing capabilities.

The uplink channel between UE~$k$ and subarray~$l$ is denoted by $\mathbf{h}_{kl} \in \mathbb{C}^{N}$, while the wireless fronthaul channel from subarray~$l$ to the CPU is represented by $\mathbf{G}_l \in \mathbb{C}^{M \times N}$. 
We assume that the uplink access transmission and the wireless fronthaul transmission take place in separate time slots.

In this work, we explicitly account for hardware impairments on both the access and fronthaul links and consider AF--based wireless fronthauling scheme as proposed in~\cite{demir2025cell}. 
Furthermore, perfect channel state information (CSI) is assumed to be available at both the subarrays and the CPU.
\vspace{-2mm}

\subsection{Uplink Data Transmission}
In the first slot of the AF wireless fronthauling scheme, the UEs send their data signals to the XL-MIMO subarrays. We denote the unit-power symbol of UE $k$ by $s_k$, i.e., $\mathbb{E}\{|s_k|^2\}=1$ and its uplink transmit power as $p_k>0$.
The received signal at subarray $l$ is modeled as
\vspace{-2mm}
\begin{align}
\mathbf{y}_l
= \sqrt{\kappa_{\mathrm{ac}}}
\sum_{k=1}^K \mathbf{h}_{kl}\sqrt{p_k} s_k
+ \boldsymbol{\eta}_{\mathrm{ac},l}
+ \mathbf{n}_{\mathrm{ac},l}, \label{eq:access}
\end{align}
where $\kappa_{\mathrm{ac}} \in (0,1]$ denotes access-side hardware quality factor, with larger values indicating weaker hardware impairment effects.
The distortion $\boldsymbol{\eta}_{\mathrm{ac},l}$ is modeled as
\vspace{-2mm}
\begin{align}
\boldsymbol{\eta}_{\mathrm{ac},l} \sim
\mathcal{CN}(\mathbf{0}, \mathbf{D}_{\mathrm{ac},l}),
\end{align}
with a covariance matrix (conditioned on the CSI) proportional to the signal power, given as \cite{massive_mimo_book}
\begin{align}
    \vect{D}_{\mathrm{ac},l} = \diag\left( (1-\kappa_{\mathrm{ac}}) \sum_{k=1}^Kp_k\vect{h}_{kl}\vect{h}_{kl}^{\Htran}\right),
\end{align}
where $\diag(\cdot)$ extracts the diagonal part of the matrix by setting off-diagonal entries to zero. The distortion is independent of the other signals and also spatially uncorrelated across antennas. The additive independent noise is denoted by $\vect{n}_{\mathrm{ac},l}\sim \CN(\vect{0},\sigma^2\vect{I}_N)$.

In the second slot, each subarray $l$ applies a fronthaul precoder $\mathbf{P}_l\in \mathbb{C}^{N\times N}$ with amplification gain $\alpha_l$. Hence, the impaired transmitted signal from subarray $l$ is given as
\begin{align}
\tilde{\mathbf{y}}_l
= \sqrt{\kappa_{\mathrm{frt}}}\alpha_l\mathbf{P}_l \mathbf{y}_l
+ \boldsymbol{\eta}_{\mathrm{frt},l}, \label{eq:transmit-fronthaul}
\end{align}
where $\kappa_{\mathrm{frt}}\in (0,1]$ is the fronthaul hardware quality factor and
$\boldsymbol{\eta}_{\mathrm{frt},l}$ is fronthaul distortion modeled as
$
\boldsymbol{\eta}_{\mathrm{frt},l}
\sim \mathcal{CN}(\mathbf{0}, \alpha_l^2\mathbf{D}_{\mathrm{frt},l})
$, where
\vspace{-2mm}
    \begin{align}
    &\vect{D}_{\mathrm{frt},l}= \diag\left( (1-\kappa_{\mathrm{frt}})\left( \sum_{k=1}^Kp_k\vect{P}_l\vect{h}_{kl}\vect{h}_{kl}^{\Htran}\vect{P}_l^{\Htran}+\sigma^2\vect{P}_l\vect{P}_l^{\Htran}\right)\right).
\end{align}
The CPU receives
\vspace{-2mm}
\begin{align}
\mathbf{y}
= \sum_{l=1}^L \mathbf{G}_l \tilde{\mathbf{y}}_l 
+ \mathbf{n}_{\mathrm{frt}}, \label{eq:cpu_received}
\end{align}
where $\mathbf{n}_{\mathrm{frt}}\sim\CN(\vect{0},\sigma^2\vect{I}_M)$ is the  additive independent noise. Next, we will consider the selection of fronthaul precoding matrix $\vect{P}_l$ at each subarray.  

\subsection{Fronthaul Precoding}

The transmit power of subarray $l$ over the fronthaul is given by
\begin{align}
    p_{\mathrm{frt},l} &=\alpha_l^2 \tr\left( \sum_{k=1}^Kp_k\vect{P}_l\vect{h}_{kl}\vect{h}_{kl}^{\Htran}\vect{P}_l^{\Htran}+\sigma^2\vect{P}_l\vect{P}_l^{\Htran}\right). \label{eq:fronthaul-power}
\end{align}

The precoding matrix $\vect{P}_l$ can be selected with an arbitrary norm, while the scalar $\alpha_l$ is used to regulate the resulting transmit power.
Following \cite{demir2025cell}, we adopt an effective fronthaul precoding strategy inspired by the optimal singular value decomposition (SVD) precoder for point-to-point MIMO transmission.
By matching the right and left singular subspaces of the precoding matrix to the access and fronthaul channels, respectively, we obtain the so-called \emph{bi-SVD precoding} scheme.

 To this end, we denote the SVDs of the fronthaul channel $\vect{G}_l$ and the access channel matrix $\vect{H}_l=[\vect{h}_{1l} \ \ldots \ \vect{h}_{Kl}]$ as
 \vspace{-2mm}
\begin{align}
\mathbf{G}_l = \mathbf{U}_{G,l} \boldsymbol{\Sigma}_{G,l} \mathbf{V}_{G,l}^{\Htran},\qquad
\mathbf{H}_l = \mathbf{U}_{H,l} \boldsymbol{\Sigma}_{H,l} \mathbf{V}_{H,l}^{\Htran}.
\end{align}
Then, the bi-SVD precoder is given as
\vspace{-2mm}
\begin{align}
\mathbf{P}_l = \mathbf{V}_{G,l}\mathbf{U}_{H,l}^{\Htran}
\end{align}
which aligns the subspaces of the access and fronthaul links, thereby improving the conditioning of the effective end-to-end channel.

In the remainder of the paper, we fix $\vect{P}_l$ according to the bi-SVD design and jointly optimize the fronthaul power control coefficients $\{\alpha_l\}$ and the UE transmit powers $\{p_k\}$ to maximize the sum rate.
 
\section{Uplink Spectral Efficiency}

Following \cite{demir2025cell}, we express the uplink SE directly in terms of the UE transmit powers $\{p_k\}$ and fronthaul power control coefficients $\{\alpha_l\}$.

We first define the effective end-to-end channel between UE $k$ and the CPU
through subarray $l$ as
\begin{align}
\mathbf{b}_{kl} \triangleq \mathbf{G}_l \mathbf{P}_l \mathbf{h}_{kl}.
\end{align}
Aggregating the contributions from all subarrays, we define the effective
end-to-end channel vector of UE $k$ as
\begin{align}
\mathbf{g}_k \triangleq \sum_{l=1}^L \alpha_l \mathbf{b}_{kl}.
\label{eq:gk_def}
\end{align}

Assuming linear combining at the CPU and treating multiuser interference and
hardware distortions as uncorrelated additive noise, the SINR-optimal
distortion-aware receive combiner for UE $k$ is given by
\begin{align}
\mathbf{v}_k^{\mathrm{opt}} \propto \mathbf{R}_k^{-1}\mathbf{g}_k ,
\label{eq:vk_opt}
\end{align}
where $\mathbf{R}_k$ is the interference-plus-distortion covariance matrix given as
\vspace{-2mm}
\begin{align}
\mathbf{R}_k
&=
\kappa_{\mathrm{frt}}\kappa_{\mathrm{ac}}
\sum_{i\neq k} p_i\, \mathbf{g}_i \mathbf{g}_i^{\Htran}
\nonumber\\
&\quad
+ \kappa_{\mathrm{frt}}
\sum_{l=1}^L \alpha_l^2\,
\mathbf{G}_l \mathbf{P}_l
\bigl(\mathbf{D}_{\mathrm{ac},l} + \sigma^2 \mathbf{I}_N\bigr)
\mathbf{P}_l^{\Htran} \mathbf{G}_l^{\Htran}
\nonumber\\
&\quad
+ \sum_{l=1}^L \alpha_l^2\,
\mathbf{G}_l \mathbf{D}_{\mathrm{frt},l} \mathbf{G}_l^{\Htran}
+ \sigma^2 \mathbf{I}_M .
\label{eq:Rk}
\end{align}

Substituting \eqref{eq:vk_opt} into the SINR expression yields the achievable
uplink SE
\begin{align}
\mathrm{SE}_k
= \log_2\!\left(1+\mathrm{SINR}_k\right),
\end{align}
with
\begin{align}
\mathrm{SINR}_k
= \kappa_{\mathrm{frt}}\kappa_{\mathrm{ac}} p_k\,
\mathbf{g}_k^{\Htran}
\mathbf{R}_k^{-1}
\mathbf{g}_k .
\label{eq:sinr_final}
\end{align}

Unlike \cite{demir2025cell}, where the uplink SE is evaluated for given transmit powers and fronthaul amplification coefficients, in this work, we explicitly target \emph{sum-rate maximization} by jointly optimizing the UE uplink transmit powers $\{p_k\}$ and the fronthaul power control coefficients $\{\alpha_l\}$. 
The resulting optimization problem is inherently non-convex due to the coupled appearance of $\{p_k\}$ and $\{\alpha_l\}$ in both the desired signal term and the interference-plus-distortion covariance matrix $\mathbf{R}_k$.
To efficiently tackle this challenge, we will develop a novel iterative algorithm inspired by the WMMSE framework, which enables alternating optimization over the transmit powers and fronthaul amplification coefficients.
This approach yields a computationally efficient and convergent solution method for joint uplink and fronthaul power control in modular XL-MIMO systems with hardware impairments.

\vspace{-2mm}

\section{Joint UE and Wireless Fronthaul Power Control}

In this section, we formulate the joint uplink and fronthaul power control problem and outline a WMMSE-based iterative solution.
\vspace{-2mm}

\subsection{Problem Formulation}

Our objective is to maximize the uplink sum SE by jointly optimizing the UE transmit powers $\{p_k\}$ and the fronthaul power control coefficients $\{\alpha_l\}$, while respecting both UE-side and fronthaul power constraints. The optimization problem is formulated as
\vspace{-2mm}
\begin{subequations}\label{eq:problem}
\begin{align}
\underset{\{p_k\}, \{\alpha_l\}}{\textrm{maximize}} \quad 
& \sum_{k=1}^K \log_2\!\left(1+\mathrm{SINR}_k\right) \label{eq:sr_problem} \\
\textrm{subject to} \quad
& 0 \le p_k \le p^{\max}_{\rm ue}, \quad \forall k, \label{eq:pk_constraint}\\
& p_{\mathrm{frt},l} \le p_{\mathrm{frt}}^{\max}, \quad \forall l, \label{eq:frt_constraint}
\end{align}
\end{subequations}
where $\mathrm{SINR}_k$ is given in \eqref{eq:sinr_final} and the fronthaul transmit power $p_{\mathrm{frt},l}$ is defined in \eqref{eq:fronthaul-power}. The $p_{\mathrm{ue}}^{\max}$ and $p_{\mathrm{frt}}^{\max}$ are the maximum allowable UE and subarray fronthaul transmit power, respectively.

Problem \eqref{eq:problem} is non-convex due to the coupled dependence of $\{p_k\}$ and $\{\alpha_l\}$ in both the desired signal term and the interference-plus-distortion covariance matrix $\mathbf{R}_k$. Direct optimization is therefore intractable.
\vspace{-2mm}

\subsection{WMMSE Reformulation}

To efficiently address \eqref{eq:sr_problem}, we adopt a WMMSE framework, building on the classical equivalence
between sum-rate maximization and weighted sum mean square error (MSE) minimization
\cite{shi2011iteratively}.

Defining $\rho_k \triangleq \kappa_{\mathrm{frt}}\kappa_{\mathrm{ac}} p_k$, the received signal can be expressed as
\vspace{-2mm}
\begin{align}
\mathbf{y} = \sqrt{\rho_k}\,\mathbf{g}_k s_k + \mathbf{z}_k,
\end{align}
where $\mathbf{z}_k$ aggregates multiuser interference, amplified distortions,
and thermal noise. Under the adopted uncorrelated distortion model,
$\mathbf{z}_k$ is uncorrelated with $s_k$ and has covariance
$\mathbb{E}\{\mathbf{z}_k\mathbf{z}_k^{\Htran}\}=\mathbf{R}_k$ given in
\eqref{eq:Rk}.

Given a linear estimate $\hat{s}_k=\mathbf{v}_k^{\Htran}\mathbf{y}$,
the resulting MSE is  
\begin{align}
e_k
&\triangleq \mathbb{E}\!\left\{ | s_k - \mathbf{v}_k^{\Htran} \mathbf{y} |^2 \right\} \nonumber\\
&= 1 - 2\Re\!\left(\sqrt{\rho_k}\,\mathbf{v}_k^{\Htran}\mathbf{g}_k\right)
+ \mathbf{v}_k^{\Htran}\!\left(\mathbf{R}_k
+ \rho_k \mathbf{g}_k\mathbf{g}_k^{\Htran}\right)\!\mathbf{v}_k .
\label{eq:ek_general}
\end{align}
For fixed $\{p_k\}$ and $\{\alpha_l\}$, minimizing \eqref{eq:ek_general}
with respect to $\mathbf{v}_k$ yields the distortion-aware linear MMSE receiver
\begin{align}
\mathbf{v}_k^{\mathrm{MMSE}}
=
\left(\mathbf{R}_k + \rho_k \mathbf{g}_k\mathbf{g}_k^{\Htran}\right)^{-1}
\sqrt{\rho_k}\,\mathbf{g}_k .
\label{eq:vk_mmse_rank1}
\end{align}

Substituting \eqref{eq:vk_mmse_rank1} into \eqref{eq:ek_general} yields the
fundamental relationship \vspace{-2mm}
\begin{align}
e_k^{\star}
= \frac{1}{1+\rho_k \mathbf{g}_k^{\Htran}\mathbf{R}_k^{-1}\mathbf{g}_k}
= \frac{1}{1+\mathrm{SINR}_k},
\end{align}
which is the key link exploited by the WMMSE reformulation.

Introducing auxiliary positive weights $\{w_k\}$, the sum-SE maximization
problem can be equivalently transformed into the following weighted sum-MSE
minimization problem:
\vspace{-2mm}
\begin{subequations}
\begin{align}
\underset{\{p_k\}, \{\alpha_l\}, \{w_k\}, \{\mathbf{v}_k\}}{\textrm{minimize}} \quad
 &\sum_{k=1}^K \left( w_k e_k - \ln (w_k) \right) \label{eq:wmmse_problem} \\
 \textrm{subject to} \quad
 &0 \le p_k \le p_{\rm ue}^{\max}, \quad \forall k, \\
& p_{\mathrm{frt},l} \le p_{\mathrm{frt}}^{\max}, \quad \forall l.
\end{align}
\end{subequations}
The proposed algorithm alternates over four variable blocks:
the receive combiners $\{\mathbf{v}_k\}$, the MSE weights $\{w_k\}$,
the UE transmit powers $\{p_k\}$, and the fronthaul power control coefficients
$\{\alpha_l\}$. Specifically, at each iteration we update
$\{\mathbf{v}_k\}$ via \eqref{eq:vk_mmse_rank1}, $\{w_k\}$ in closed form, and
then solve two convex subproblems to update $\{p_k\}$ and $\{\alpha_l\}$ while
keeping the other variables fixed. For fixed $\{p_k\}$, $\{\alpha_l\}$, and
$\{\mathbf{v}_k\}$, the optimal weights are
\begin{align}
w_k^{\star} = e_k^{-1}, \quad \forall k.
\end{align}

\subsection{Numerical Updates of UE Powers and Fronthaul Amplification Coefficients}

In this subsection, we describe the numerical updates of the UE transmit powers
and fronthaul amplification coefficients within the WMMSE iterations.
Throughout this subsection, we fix $\{\mathbf v_k\}$ and $\{w_k\}$ and update
$\{p_k\}$ and $\{\alpha_l\}$ by solving two convex subproblems.

\subsubsection{UE Power Update via a Convex Program}

Consider the WMMSE objective in \eqref{eq:wmmse_problem} for fixed
$\{\alpha_l\}$, $\{\mathbf v_k\}$, and $\{w_k\}$.
Define $q_k\triangleq \sqrt{p_k}\ge 0$ and let $c\triangleq \kappa_{\mathrm{frt}}\kappa_{\mathrm{ac}}$.
For fixed $\{\alpha_l\}$, the weighted MSE term can be written as a convex
quadratic function of $\mathbf q\triangleq [q_1,\ldots,q_K]^{\Ttran}$:
\begin{align}
\sum_{k=1}^K w_k e_k
=
\text{const}
+\sum_{k=1}^K \left(A_k\, q_k^2 - 2 B_k\, q_k\right),
\label{eq:wmmse_quad_q_vec}
\end{align}
where $A_k\ge 0$ and $B_k$ are given by
\begin{align}
A_k
&\triangleq
w_k c
\left|
\mathbf v_k^{\Htran}\mathbf g_k
\right|^2
+
\sum_{j=1}^K w_j\,\phi_{j,k}
\label{eq:Ak_alpha_full}\\
B_k
&\triangleq
w_k \sqrt{c}\,
\Re\!\left(
\mathbf v_k^{\Htran}\mathbf g_k
\right)
=
w_k \sqrt{c}\,
\Re\!\left(
\mathbf v_k^{\Htran}
\sum_{l=1}^L \alpha_l \mathbf b_{kl}
\right),
\label{eq:Bk_alpha_full}
\end{align}
with
\begin{align}
\phi_{j,k}
\triangleq
\mathbf v_j^{\Htran}
\Big(
\mathbf C^{(\mathrm{int})}_{j,k}
+
\mathbf C^{(\mathrm{ac})}_{k}
+
\mathbf C^{(\mathrm{frt})}_{k}
\Big)
\mathbf v_j ,
\label{eq:phi_jk_alpha_full}
\end{align}
where
\vspace{-2mm}
\begin{align}
&\mathbf C^{(\mathrm{int})}_{j,k}
\triangleq
c\,\mathbbm{1}_{\{k\neq j\}}
\vect{g}_{k}\vect{g}_k^{\Htran},
\label{eq:Cint_jk_alpha} \\
&\mathbf C^{(\mathrm{ac})}_{k}
\triangleq
\kappa_{\mathrm{frt}}
\sum_{l=1}^L \alpha_l^2\,
\mathbf G_l\mathbf P_l\,
\diag\!\Big(
(1-\kappa_{\mathrm{ac}})
\mathbf h_{kl}\mathbf h_{kl}^{\Htran}
\Big)\,
\mathbf P_l^{\Htran}\mathbf G_l^{\Htran},
\label{eq:Cac_jk_alpha} \\
&\mathbf C^{(\mathrm{frt})}_{k}
\triangleq
\sum_{l=1}^L \alpha_l^2\,
\mathbf G_l\,
\diag\!\Big(
(1-\kappa_{\mathrm{frt}})
\mathbf P_l\mathbf h_{kl}\mathbf h_{kl}^{\Htran}\mathbf P_l^{\Htran}
\Big)\,
\mathbf G_l^{\Htran}.
\label{eq:Cfrt_jk_alpha}
\end{align}
Moreover, each fronthaul transmit power constraint \eqref{eq:fronthaul-power} can be written as
\vspace{-2mm}
\begin{align}
p_{\mathrm{frt},l}
=
\alpha_l^2\left(\sum_{k=1}^K \xi_{l,k}\, q_k^2 + \sigma^2 \tr(\mathbf P_l\mathbf P_l^{\Htran})\right)
\le p_{\mathrm{frt}}^{\max},
\label{eq:frt_q_constraint}
\end{align}
where $\xi_{l,k}\triangleq \tr(\mathbf P_l\mathbf h_{kl}\mathbf h_{kl}^{\Htran}\mathbf P_l^{\Htran})
=\|\mathbf P_l\mathbf h_{kl}\|^2$.

Hence, updating $\mathbf q$ reduces to the convex quadratically-constrained quadratic problem (QCQP)
\begin{subequations}\label{eq:q_qcqp}
\begin{align}
\underset{\mathbf q}{\textrm{minimize}} \quad
& \sum_{k=1}^K \left(A_k\, q_k^2 - 2 B_k\, q_k\right) \\
\textrm{subject to} \quad
& 0 \le q_k \le \sqrt{p_{\rm ue}^{\max}}, \quad \forall k,\\
& \hspace{-8mm}\alpha_l^2\left(\sum_{k=1}^K \xi_{l,k}\, q_k^2 + \sigma^2 \tr(\mathbf P_l\mathbf P_l^{\Htran})\right)
\le p_{\mathrm{frt}}^{\max}, \quad \forall l.
\end{align}
\end{subequations}
Problem \eqref{eq:q_qcqp} is convex and can be efficiently solved using standard numerical solvers.
The UE powers are then updated as $p_k \leftarrow q_k^2$.

\subsubsection{Fronthaul Amplification Update via a Convex Quadratic Program}

Next, for fixed $\mathbf q$ (equivalently, fixed $\{p_k\}$), $\{\mathbf v_k\}$, and $\{w_k\}$,
we update the fronthaul amplification coefficients $\boldsymbol\alpha=[\alpha_1,\ldots,\alpha_L]^{\Ttran}$.

We define for each pair $(k,i)$
\begin{align}
\mathbf t_{k,i}\triangleq
\begin{bmatrix}
\mathbf v_k^{\Htran}\mathbf b_{i1} & \mathbf v_k^{\Htran}\mathbf b_{i2} & \cdots &
\mathbf v_k^{\Htran}\mathbf b_{iL}
\end{bmatrix}^{\Ttran}\in \mathbb{C}^{L},
\end{align}
so that $\mathbf v_k^{\Htran}\mathbf g_i=\boldsymbol\alpha^{\Ttran}\mathbf t_{k,i}$. Then, \vspace{-2mm}
\begin{align}
\mathbf v_k^{\Htran}\mathbf R_k\mathbf v_k
&=
c\sum_{i\neq k} p_i\,\bigl|\boldsymbol\alpha^{\Ttran}\mathbf t_{k,i}\bigr|^2
+\sum_{l=1}^L \alpha_l^2\,\delta_{k,l}
+\sigma^2\|\mathbf v_k\|^2,
\label{eq:vkRkvk_coherent}
\end{align}
where \vspace{-2mm}
\begin{align}
&\delta_{k,l}\triangleq \nonumber \\
&\mathbf v_k^{\Htran}\Big(
\kappa_{\mathrm{frt}}\mathbf G_l\mathbf P_l(\mathbf D_{\mathrm{ac},l}+\sigma^2\mathbf I_N)\mathbf P_l^{\Htran}\mathbf G_l^{\Htran}
+\mathbf G_l\mathbf D_{\mathrm{frt},l}\mathbf G_l^{\Htran}
\Big)\mathbf v_k \ \ge 0 .
\label{eq:delta_kl_def}
\end{align}

Collecting the $\boldsymbol\alpha$-dependent terms in $\sum_k w_k e_k$, we obtain
\begin{align}
\sum_{k=1}^K w_k e_k
=
\text{const}
+ \boldsymbol\alpha^{\Htran}\mathbf H\,\boldsymbol\alpha
-2\,\Re(\mathbf f^{\Htran}\boldsymbol\alpha),
\label{eq:wmmse_alpha_quad_coherent}
\end{align}
with
\begin{align}
\mathbf f
&\triangleq
\sum_{k=1}^K w_k\sqrt{c p_k}\,\vect{t}_{k,k},
\\
\mathbf H
&\triangleq
c\sum_{k=1}^K w_k \sum_{i=1}^Kp_i\,\vect{t}_{k,i}\vect{t}_{k,i}^{\Htran} \nonumber\\
&\quad\;+\;
\diag\!\left(\sum_{k=1}^K w_k\delta_{k,1},\ldots,\sum_{k=1}^K w_k\delta_{k,L}\right).
\label{eq:H_def_coherent}
\end{align}

The fronthaul constraints for fixed $\{p_k\}$ are
\begin{align}
\alpha_l^2\,\psi_l \le p_{\mathrm{frt}}^{\max}, \qquad \forall l,
\end{align}
where \vspace{-2mm}
\begin{align}
\psi_l \triangleq
\tr\!\left(\sum_{k=1}^K p_k\,\mathbf P_l\mathbf h_{kl}\mathbf h_{kl}^{\Htran}\mathbf P_l^{\Htran}
+\sigma^2\mathbf P_l\mathbf P_l^{\Htran}\right).
\end{align}
Hence, the $\boldsymbol\alpha$-update is the convex quadratic problem
\begin{subequations}\label{eq:alpha_qcqp_coherent}
\begin{align}
\underset{\boldsymbol\alpha}{\textrm{minimize}} \quad
& \boldsymbol\alpha^{\Htran}\mathbf H\,\boldsymbol\alpha
-2\,\Re(\mathbf f^{\Htran}\boldsymbol\alpha) \\
\textrm{subject to} \quad
& 0\leq \alpha_l\leq \ \sqrt{\frac{p_{\mathrm{frt}}^{\max}}{ \psi_l}}, \quad \forall l.
\end{align}
\end{subequations}
Problem \eqref{eq:alpha_qcqp_coherent} is convex since $\mathbf H\succeq \mathbf 0$ and the constraints are affine.
It can be efficiently solved using standard numerical solvers.

The above steps are iterated until convergence. Since each iteration monotonically decreases the WMMSE objective, the proposed algorithm is guaranteed to converge \cite{shi2011iteratively}.
\vspace{-2mm}

\section{Numerical Results}

In this section, we evaluate the performance of the proposed modular XL-MIMO architecture with AF-based wireless fronthaul and quantify the gains obtained from joint uplink power and fronthaul amplification optimization. 
We compare three cases: i) an unoptimized baseline with fixed UE powers and fixed fronthaul amplification, ii) joint optimization of UE powers and fronthaul amplification coefficients, and iii) an ideal-hardware benchmark where hardware impairments are disabled. 

We adopt the path-loss and local scattering framework from \cite{cell-free-book}. 
The simulation considers a square coverage area of $200\times200$\,m$^2$, where $K=8$ UEs are independently and uniformly dropped in the square whose corners are $(0,0)$\,m and $(200,200)$\,m. 
The $L=64$ modular subarrays (with $N=4$ antennas each) form an $8\times 8$ structure with $10$\,m vertical and horizontal spacing: they are placed along a line at $y=200$\,m with horizontal offsets around $x=100$\,m, while their heights are $z\in\{10,20,\ldots,80\}$\,m. 
The CPU is located at $(x,y)=(100,200)$\,m and height $z=90$\,m, and the fronthaul distance includes the vertical separation between each subarray and the CPU. The access links follow correlated Rayleigh fading generated from spatial correlation matrices obtained via the local scattering model with angular spreads $\text{ASD}_\varphi=15^{\circ}$ and $\text{ASD}_\theta=15^{\circ}$, whereas the fronthaul channels follow correlated Rician fading with a K-factor of $10$\,dB. 
The vertical distance between the arrays and the UEs is set to $10$\,m.

The communication bandwidth is $50$\,MHz and the noise figure is $3$\,dB, which results in a noise power of approximately $-94$\,dBm. 
Unless otherwise stated, the hardware quality factors are $\kappa_{\rm ac}=\kappa_{\rm frt}=0.95$. 
Each UE transmit power satisfies $0\le p_k\le 0.2$\,W, and each subarray enforces a fronthaul transmit power constraint corresponding to a calibrated $10$\,W reference level. 
Fronthaul precoding is implemented using a bi-SVD construction that aligns the access and fronthaul spatial modes.

We report the cumulative distribution function (CDF) of the per-UE SE in Figs.~1 and~2 for $M=12$ and $M=24$, respectively. 
The case labeled ``Perfect hardware'' represents the ideal SE achievable with two-phase processing under the considered wireless fronthauling scheme. 
We also present the SE obtained with the proposed joint resource allocation via the WMMSE algorithm and a fixed-power baseline that employs maximum uplink power together with maximum amplification gain at the wireless fronthaul. 
The red curves correspond to the case where only the access links are impaired while the fronthaul hardware is ideal, whereas the black curves represent the scenario where both the access and fronthaul links are impaired.

By comparing the two figures, we first observe that increasing the number of CPU antennas shifts all curves to the right, indicating improved SE performance. 
Moreover, the degradation caused by access-side impairments alone is already substantial, while the additional impairment on the fronthaul introduces a comparatively smaller but non-negligible loss. 
When the CPU employs fewer antennas, the proposed WMMSE algorithm provides a more pronounced improvement in per-UE SE compared to the fixed-power strategy; however, this performance gap becomes smaller as $M$ increases.

\begin{figure}[t] 
    \centering
        \includegraphics[width=0.46\textwidth, trim=0.2cm 0cm 1cm 0.2cm, clip]{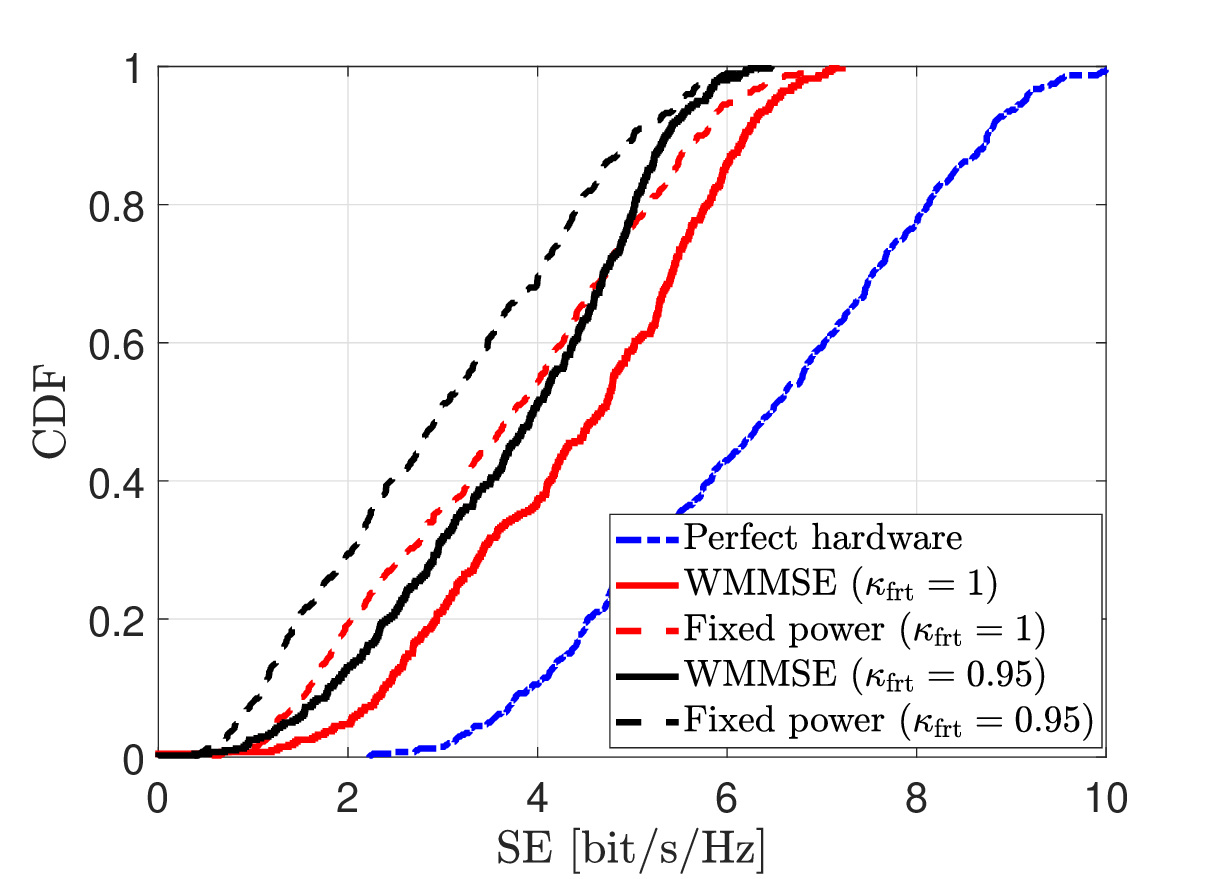} 
                \vspace{-2mm}
        \caption{The CDF of the uplink SE for $M=12$.}
        \label{fig:1}
        \vspace{-7mm}
 \end{figure}

 \begin{figure}[t] 
    \centering
        \includegraphics[width=0.46\textwidth, trim=0.2cm 0cm 1cm 0.2cm, clip]{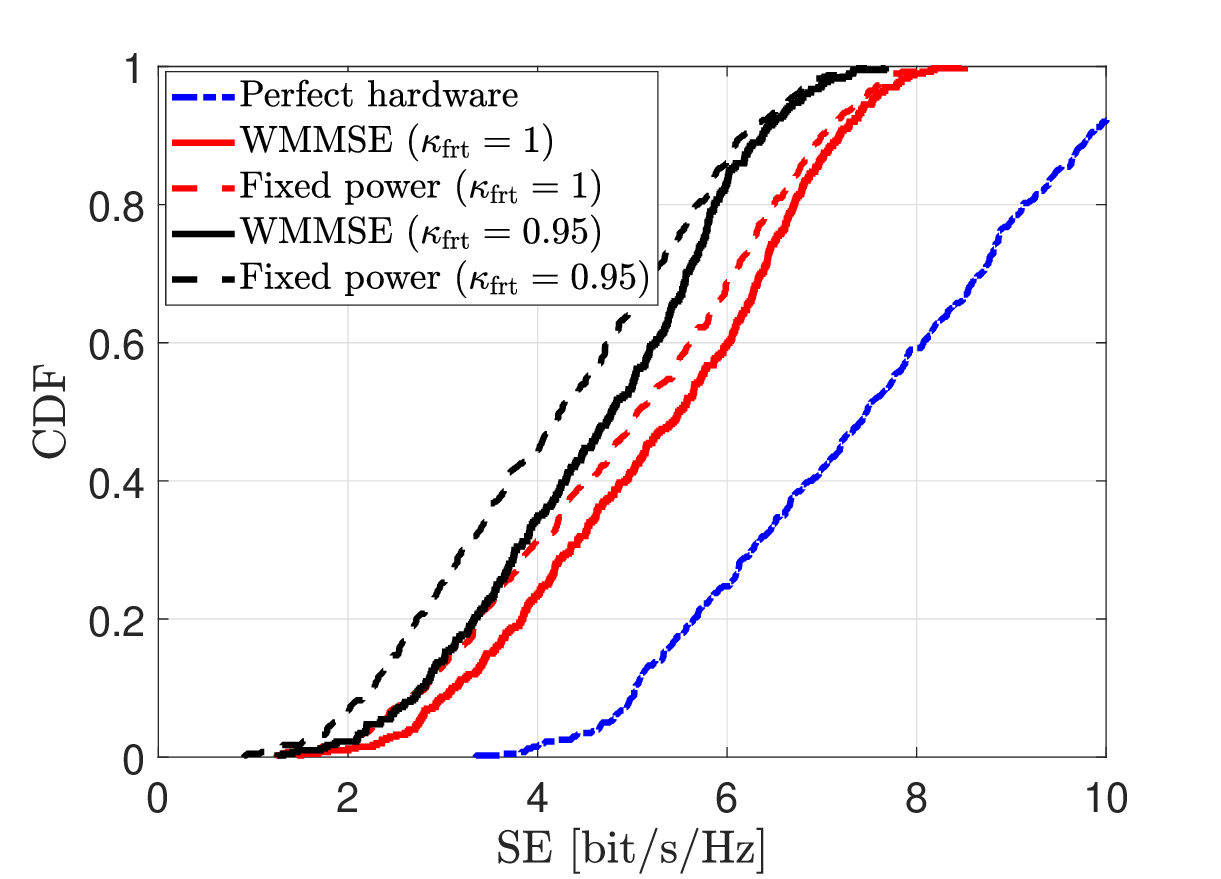}
                \vspace{-2mm}
        \caption{The CDF of the uplink SE for $M=24$.}
                \vspace{-7mm}
        \label{fig:1}
 \end{figure}

\section{Conclusions}
This paper investigated distributed two-phase processing for modular XL-MIMO systems with AF-based wireless fronthaul under hardware impairments. 
By jointly optimizing UE transmit powers and fronthaul amplification coefficients using a WMMSE-based framework, we derived an efficient algorithm that accounts for distortion-aware signal processing and practical power constraints. 
The results show that adaptive joint optimization provides noticeable gains over fixed-power strategies, especially when the number of CPU antennas is limited, while access-side impairments remain the dominant source of performance degradation. 
Overall, the proposed framework highlights the potential of modular XL-MIMO with wireless fronthaul as a scalable architecture for future distributed large-aperture systems.

\bibliographystyle{IEEEtran}
\bibliography{IEEEabrv,refs}

\end{document}